\documentclass[manuscript]{aastex}

\shorttitle{Parameters for ten LMC SCs}
\shortauthors{Ahumada et al.}

\begin{document}
\title{Determination of reddening and age for ten Large Magellanic Cloud star clusters from integrated spectroscopy}

\author{Andrea V. Ahumada}
\affil{Observatorio Astron\'omico, Universidad Nacional de C\'ordoba,
Laprida 854, 5000, C\'ordoba, Argentina.\\
Consejo Nacional de Investigaciones Cient\'{\i}ıficas y T\'ecnicas (CONICET), Argentina.}
\email{andrea@oac.uncor.edu}

\author{Luis R. Vega}
\affil{Observatorio Astron\'omico, Universidad Nacional de C\'ordoba,
Laprida 854, 5000, C\'ordoba, Argentina\\
IATE, Consejo Nacional de Investigaciones Cient\'{\i}ıficas y T\'ecnicas (CONICET), Argentina.\\
Secretar\'{\i}a de Ciencia y T\'ecnica, Universidad Nacional de C\'ordoba, C\'ordoba, Argentina.}
\email{luis@oac.unc.edu.ar}

\author{Juan J. Clari\'a}
\affil{Observatorio Astron\'omico, Universidad Nacional de C\'ordoba,
Laprida 854, 5000, C\'ordoba, Argentina\\
Consejo Nacional de Investigaciones Cient\'{\i}ıficas y T\'ecnicas (CONICET), Argentina.}
\email{claria@oac.uncor.edu}

\author{M\'onica A. Oddone\altaffilmark{1}}
\affil{Observatorio Astron\'omico, Universidad Nacional de C\'ordoba,
Laprida 854, 5000, C\'ordoba, Argentina}
\email{mao@oac.uncor.edu}

\and
\author{Tali Palma\altaffilmark{1}}
\affil{Millennium Institute of Astrophysics, Chile. \\
Instituto de Astrof\'{\i}sica, Pontificia Universidad Cat\'olica de Chile, Av. Vicu\~na Mackenna 4860, 782-0436 Macul, Santiago, Chile.\\
Observatorio Astron\'omico, Universidad Nacional de C\'ordoba,
Laprida 854, 5000, C\'ordoba, Argentina}
\email{tpalma@astro.puc.cl}

\altaffiltext{1}{Visiting Astronomer, Complejo Astron\'omico El Leoncito operated under agreement between the Consejo Nacional de Investigaciones Cient\'{\i}ficas y T\'ecnicas de la Rep\'ublica Argentina and the National Universities of La Plata, C\'ordoba and San Juan.}

\begin{abstract}
We present flux-calibrated integrated spectra in the optical range (3700-6800 \AA) obtained at Complejo Astron\'omico El Leoncito (CASLEO, Argentina) for a sample of 10 concentrated star clusters belonging to the Large Magellanic Cloud (LMC). No previous data exist for two of these objects (SL\,142 and SL\,624), while most of the remaining clusters have been only poorly studied. We derive simultaneously foreground $E(B-V)$ reddening values and ages for the cluster sample  by comparing their integrated spectra with template LMC cluster spectra and with two different sets of simple stellar population models. Cluster reddening values and ages are also derived from both available interstellar extinction maps and by using diagnostic diagrams involving the sum of equivalent widths of some selected spectral features and their calibrations with age, respectively. For the studied sample, we derive ages between 1 Myr and 240 Myr. In an effort to create a spectral library at the LMC metallicity level with several clusters per age range, the cluster sample here presented stands out as a useful complement to previous ones.
\end{abstract}





\section{Introduction}

Star clusters (SCs) are groups of coeval stars that formed from the same molecular cloud so they can be considered as building blocks of galaxies. This is the reason why the study of rich extragalactic SCs provides valuable information about the star formation and chemical histories of the host galaxies. However, the current knowledge of both the stellar formation processes and the chemical evolution of distant galaxies is still far from being complete. Even for the galaxies in the Local Group, there is a very limited understanding of these processes. In this context, the SCs of the Large Magellanic Cloud (LMC), on account on their proximity, richness and variety can facilitate our comprehension of the chemical enrichment and star formation history of this galaxy as a whole \citep{ric01,bau13}. In fact, LMC SCs can be resolved in their individual members and can therefore be studied from their color-magnitude diagrams. Although the LMC SCs recently estimated number is approximately 3100 \citep{bic08}, this number may be significantly larger if emission-free associations and objects related to emission nebulae are included \citep{bic99}.\\

Integrated spectroscopy is a powerful technique to study SCs in galaxies \citep{bic88,sch05}. In particular, template spectra are very useful in deriving individual clusters parameters not only in our Galaxy (e.g., Ahumada et al. 2007) but also in the Large and Small Magellanic Clouds (e.g., Santos et al. 2006; Piatti et al. 2005 ; Dias et al. 2010) and even in distant galaxies (e.g., Jablonka et al. 1998; Trancho et al. 2007). Integrated spectra in the visible spectral range provide good age predictions when compared with high resolution computational models \citep{asa13}.\\

One of the goals of the present study is to collect and analize a sample of spectra of LMC SCs  with the aim of studying the integrated light properties of such metal-deficient SCs, deriving their fundamental parameters (reddening and age) and making them available as template spectra to complement previous sub-solar metallicity libraries. The present study is part of an ongoing project of integrated spectroscopy of concentrated LMC SCs which is being carried out at the Observatorio Astron\'omico de la Universidad Nacional de C\'ordoba (Argentina). We have already reported results based on integrated spectra for 95 concentrated LMC SCs younger than 1 Gyr \citep{san06,tal06,tal09,pal08,ahu11,odd12,min12,min13,min14}. These studies have not only contributed to the individual characterization of all these clusters but have also allowed to generate new template spectra at the metallicity level of the LMC (e.g., Minniti et al. 2014).\\
 
\begin{figure}
\plotone{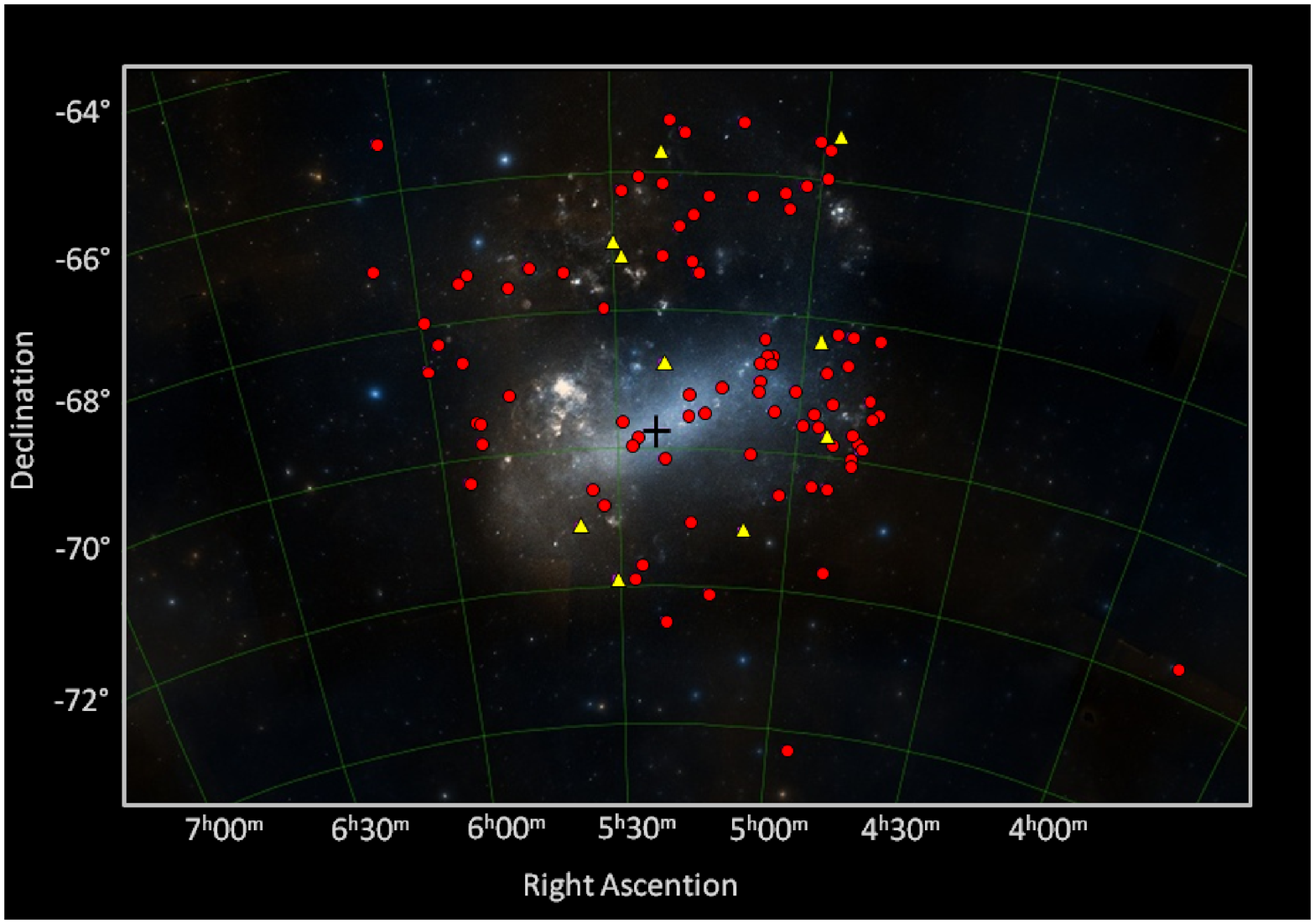}
\caption{Spatial distribution of the studied SCs superimposed on the LMC image. Red filled circles represent previous studied SCs and yellow filled triangles represent the present sample. The optical center of the LMC is represented with a cross. \label{fig1}}
\end{figure}

We present here flux-calibrated integrated spectra in the optical range for ten unstudied or poorly studied LMC SCs, with types between 0 and III in the sequence defined by Searle et al. (1980, hereafter SWB80). These SWB80 types correspond to SCs younger than 200 Myr. The obtained spectra are used to derive reddening and age for the cluster sample. We plot in Fig.~\ref{fig1} the projected spatial distribution for all the SCs we have observed up to now through integrated spectroscopy. Yellow filled triangles represent clusters studied in the current work. As the LMC center, we adopted the position of NGC 1928 ($\alpha_{2000}$ = 5h20m57s, $\delta_{2000}$ = -69$^{\circ}$28'41''), which is indicated with a cross in Fig. 1 .\\

In Section 2, we present the cluster sample and describe details of the observations and the data reduction procedure. The methods employed to derive reddening and age values, as well as the measurements of equivalent widths obtained for different absorption features are described in Section 3. A discussion of the results found for the individual clusters is presented in Section 4, and the final conclusions are given in Section 5. \\

\section{Observations and reductions}


We selected from the catalogue of UBV photometry of SCs and stellar associations of Bica et al. (1996, hereafter B96) ten concentrated and high brightness SCs located in the inner disc and in the outer regions of the LMC (Fig. 1). As far as we aware, no previous data exist for two of these objects (SL\,142 and SL\,624) and most of the remaining clusters have been only poorly studied. The observed clusters are listed in Table~\ref{tbl-1}, wherein designations from different catalogues are provided. Equatorial and Galactic coordinates for the present cluster sample are given in Table~\ref{tbl-1}, together with cluster averaged diameters and SWB80 types reported by B96. \\

The spectroscopic observations took place at Complejo Astron\'omico El Leoncito (CASLEO) in San Juan (Argentina), using the ``{\it Jorge Sahade}'' 2.15 m telescope during two observation runs in 2011 and 2013. Table~\ref{tbl-2} shows the log of the observations with dates, exposure times, and the resulting total signal-to-noise ratio (S/N) of the spectra. As in previous works (see, e.g., Minniti et al. 2014), we employed a CCD camera attached to the REOSC spectrograph in the simple mode. The detector was a Tektronics chip of 1024$\times$1024 pixels of size 24$\mu$$\times$24$\mu$. The slit was set in the east-west direction and the observations were performed by scanning the slit across the objects in the north-south direction to get a proper sampling of cluster stars. A grating of 300 grooves/mm was used, which produced an average dispersion in the observed region of $\approx$ 3.46 \AA/pixel. The useful spectral coverage was 3700-6800 \AA. The length of the slit, corresponding to 4.7' on the sky, enabled us to sample regions of background sky. The slit width was 400 $\mu$m, thus providing a resolution of 14 \AA, as measured from the full width at half maximum in the Cu-Ar-Ne lines of the comparison lamps. Standard stars taken from the list of \citet{sto83} were observed for flux calibrations. Bias, darks, and dome and twilight sky flats were obtained and employed in the reductions.\\

The reduction of the spectra was carried out with the Image Reduction and Analysis Facility ({\sc iraf})\footnote{{\sc iraf} is distributed by the National Optical Astronomy Observatory, which is operated by the Association of Universities for Research in Astronomy (AURA) under a cooperative agreement with the National Science Foundation. } software package following standard procedures at the Observatorio Astron\'omico de la Universidad Nacional de C\'ordoba (Argentina). The spectra were extracted along the slit depending on cluster size and available flux. Background sky subtractions were performed using pixel rows from the same frame. This was done after removing cosmic rays from the background sky regions, taking care that no significant background sky residuals were present on the resulting spectra. Wavelength calibrations were accomplished with a Cu-Ar-Ne lamp with exposures following that of the object or standard star. Atmospheric extinction corrections according to the site coefficients given by \citet{min89} as well as flux calibrations were then applied. We combined the individual spectra obtained for each cluster in order to get a final spectrum with an improved S/N. Finally, these spectra were normalized at approximately 4020 \AA\, by avoiding spectral lines that could eventually be present. Figure 2 shows the flux-calibrated integrated spectra of the cluster sample. Spectra are in relative flux units. Constants have been added to the spectra for comparison purposes, except for the bottom one. Note that whereas some amount of field star contamination is expected to be found in the observed spectra, only bright field stars could significantly affect them. No bright stars seem to be present in the fields of the observed clusters.

\section{Derivation of reddening and age}

We obtained ages and reddening values for the cluster sample using two different methods. On the one hand, applying the template match procedure that consists of comparing the line strengths and continuum distribution of the observed cluster spectra with those of template spectra with known properties. On the other hand, comparing the observed integrated spectra with two sets of simple stellar population (SSP) models. These two methods are briefly described below.

\subsection{{\bf Template matching method}}
To derive reddening and age values, we followed the steps described in Santos et al. (2006) and \citet{min14}. In brief, we first used the equivalent width (EW) method which allowed us to estimate a preliminary age range value. This method implies using EWs of selected spectral features and their sums, along with their calibrations with age from the works of Bica \& Alloin (1986) and Santos \& Piatti (2004, hereafter SP04). Taking into account the first preliminary age range value estimated, we then applied the template matching method which consists in comparing the observed spectra to template spectra with well-determined ages (e.g., Ahumada et al. 2002, Minniti et al. 2014). Foreground $E(B-V)$ reddening values were also determined through the template matching method. Only for comparison purposes, we have also determined $E(B-V)$ color excesses by interpolation between the extinction maps of Burstein \& Heiles (1982, hereafter BH82). We preferred not to take into account the interstellar extinction maps of \citet{sch98} because they appear to be basically saturated towards the LMC disk. 

The EWs, in Angstrom units (\AA), measured for the first four H Balmer lines, H\,Ca\,II and K\,Ca\,II, G\,band (CH) and Mg\,I (5167+5173+5184 \AA) are shown in Table~\ref{tbl-3}. We also included in this table the sum of the EWs of the three Balmer lines H$\beta$, H$\gamma$ and H$\delta$ (S$_h$), as well as the sum of the EWs of the metallic lines K\,Ca\,II, G\,band (CH) and Mg\,I (S$_m$). Typical errors of approximately 10\% on individual EW measurements were the result of tracing slightly different continua.  Based on SP04's study, \citet{san06} obtained two equations to estimate cluster ages from the sum of the EWs. We used their equation (1), based on S$_m$, to get a first age estimate, and then we used their equation (2), based on S$_h$, for a second age estimate. We also employed SP04's diagnostic diagrams (DDs) involving S$_h$ and S$_m$, which are useful for the discrimination of old, intermediate-age and young clusters. The resulting ages are listed in Table~\ref{tbl-4}. Note that, in general terms, the age ranges derived using S$_m$ and S$_h$ are fairly consistent. Balmer indices of some very young clusters that show spectral features in emission cannot be used to estimate cluster ages. Based on the first age estimates, we selected among the available templates those to be compared with the observed spectrum. We decided to use Santos et al. (1995, hereafter S95) template spectral library because it represents the blue-violet integrated spectral evolution of the young Magellanic Cloud star clusters. This library, however, includes some templates covering a limited spectral range between 3500 and 4700 \AA\, and is valid only for clusters younger than approximately 170 Myr. We should to make it clear that these templates were corrected for foreground reddening using $E(B-V)$ = 0.06 (S95).

The final age determination was reached by varying reddening and templates until the best match was obtained between the chosen template and the observed spectrum. To perform reddening corrections when applying the template matching method, we used the normal reddening law $A_{\lambda} = 0.65A_V(1/{\lambda}-0.35)$ \citep{sea79}, the relation A$_v$ = 3.0$E(B-V)$, and the Fast Integrated Spectra Analyzer (FISA, Ben\'{\i}tez-Llambay et al. 2012, hereafter BCP12) algorithm. This software permits a fast and reasonably accurate determination of a cluster age and reddening using its integrated spectrum. As explained by BCP12, the determination of the most adequate template spectrum-reddening combination is performed by minimizing the $\chi^2_{jk}$ function, i.e., BCP12's equation (16).

 The results obtained after applying the template matching method are presented in Table~\ref{tbl-5}, wherein columns (3)-(4) provide the corresponding age and $E(B-V)$ color excesses. Only for comparison purposes, we list in column (2) cluster $E(B-V)$ reddening values estimated by interpolation using the extinction maps of BH82. These values should probably be considered as the lower limit for cluster reddening. The best fits of the reddening-corrected spectra with templates from S95 are presented in Figs. 3-4. Residual fluxes computed as (F$_{cluster}$ - F$_{template}$)/F$_{cluster}$ are shown in these figures. We also indicate in Figs. 3-4 the name and age range of the selected template for each cluster.

\subsection{{\bf Full spectrum fitting}}

 We have also fitted our observed spectra through the spectral synthesis
technique. This method has proven to be a powerful tool for fitting a
given observed spectrum ($O_\lambda$) with a model ($M_\lambda$) that
best matches the observed one. As evolutionary synthetic spectra of SSPs are recently available, some fundamental parameters of stellar clusters, such as reddening, age, and metallicity can be directly obtained by means of the above mentioned technique. To achieve this goal, we chose to use the {\sc starlight} code \citep{CidF05}, which incorporates a set of SSP models of different ages and
metallicities (a ``base''). The base spectra are linearly combined in different proportions to obtain the resulting spectrum $M_\lambda$ which
best fits $O_\lambda$. Since the fits are performed by minimizing $\chi^2 =
\sum_{\lambda} [(O_\lambda - M_\lambda)/O_\lambda]^2 \times \omega_\lambda^2$,  where $\omega_\lambda$ is the weight given to the fits, the spectral regions that are not wanted to be modeled (e.g. emission lines or cosmic rays) are masked out by chosing $\omega_\lambda$ = 0. This minimization yields the best population mixture in the parameter space.


For our purposes, we selected two bases of SSPs: the first one consisting of Bruzual \& Charlot (2003, hereafter BC03) SSPs with 15 different ages ranging between 1 and 500 Myr and with the same metallicity level as the LMC ($Z/Z_{\odot}$ = 0.4). This base, hereafter ``base A'', was drawn from a more general set of 150 SSPs with 25 ages and 6 metallicities. We opted to constrain the metallicity to that of the LMC and not to consider ages older than 1 Gyr, since it is not relevant from the astrophysical point of view when we deal with LMC cluster spectra. The second base of synthetic spectra taken from BaSTI database \citep{per09}, hereafter ``base B'', includes 15 spectra with $Z/Z_{\odot}$ = 0.4, with ages ranging from 40 to 500 Myr.


It should be mentioned that although BaSTI includes both high and low resolution spectra with 11 different  metallicities and ages up to 15 Gyr, we decided to use high resolution spectra. \citet{kw06} derived a mean present-day metallicity of [Fe/H] = -0.34 for the LMC, in good agreement with previous metallicity determinations of the young LMC population. This metallicity matches better the models for Z = 0.008 (or $Z/Z_{\odot}$ = 0.4), which were now used for the age and reddening determination.

The corresponding fits are shown in Figs. 5-8, along with the
most probable combination of SSPs (right panels), i.e., the ``population vector''
formed by non-zero contributions. In general, a stellar cluster is
represented by only one component which gives 100\% of the
flux. In some cases, however, a multicomponent fit is needed due to
the coarseness of the parameters in the base, possible
contamination of foreground and background stars and the low S/N ratio, as
was pointed out by \citet{D10}. There is still another source of uncertainty in the determination of the parameters of a cluster: the lack of some SSPs in the base, known as ``template mismatch''. The right panels in Figs. 5-8 show that for most cases just one SSP is needed, while in others the population vector consists of more components. Mean ages calculated over the population vector of each fit are shown beside each vertical line. The {\sc starlight} code also computes the internal reddening of each cluster. Since the flux-calibrated spectra were also corrected for a Galactic mean extinction of $A_v$ = 0.15, any ``$E(B-V)$'' value yielded by the code will reflect an ``extra'' interstellar extinction. While in some cases this ``extra'' reddening could be due to the fact that we are using a mean Galactic reddening value to correct our spectra, in some fits the values of ``$E(B-V)$'' shown here could denote possible internal reddening of the cluster inside the LMC. In Table~\ref{tbl-5}, we list the resulting ages and $E(B-V)$ values. {\sc starlight} does not provide any kind of errors in the parameters of the models. For this reason, \citet{CidF10} analyzed the errors involved after performing simulations on the fits of star cluster spectra and derived an uncertainty of $\le$ 0.03 for $E(B-V)$ ($\le$ 0.1 for $A_V$) and $\le$ 0.1 dex for log age. Taking into account their results and the fact that our observed spectra are of comparatively high S/N ratio ($\ge$ 28) which imply more reliable determinations of the parameters, we conservatively adopted an error of 0.1 dex in the log age determinations for both simple and multiple population fits.

Figs. 5-8 show that our fits are of good quality since the residual fluxes are small in all the fittings. For SL\,106, SL\,428, SL\,498 and SL\,516, however, no reliable fits were obtained using base B, probably because these are very young clusters and, as mentioned before, BaSTI does not present SSPs for age values lower than 40 Myr. Some of the spectra belong to young stellar clusters. As these clusters are probably associated to ionized gas, their spectra may exhibit emission lines. Since the {\sc starlight} code deals with stellar populations, it only fits the continuum and absorption lines. As a result, the emission line features as well as possible cosmic rays are masked in the fits, and the residuals could show ionized gas emission in those clusters. Global $\chi^2$ values computed over non-masked spectral regions are shown for BC03 and BaSTI models. We see in Fig. 9 that the quality of the spectra is statistically better for BC03 models for which $\chi^2$ values are systematically lower.




\section{Discussion on individual clusters}

We reported in \citet{odd12} only preliminary results for 5 of the 10 clusters here analized. These results may be considered a first approach to the estimation of reddening and age values. As mentioned in the introduction, there are no previous data for SL\,142 and SL\,624 so that foreground $E(B-V)$ values and ages of these two clusters are determined here for the first time. The results obtained for each of the 10 SCs are briefly discussed below.  Table~\ref{tbl-5} presents ages and reddening values obtained after fittings with S95 templates and also with SSPs bases A and B. We also list the finally adopted parameters derived by assigning different weights to the obtained values in the fits with S95 and with the SSPs of BC03. BaSTI values are included in Table~\ref{tbl-5} only as a reference, given the reasons stated in Section 3 (see Fig. 9). The errors included in the finally adopted parameters represent our estimations performed taking into account the different determinations as well as the different weights assigned to them. 

\subsection{SL\,106}
This is a young and compact group of stars classified by B96 as belonging to SWB\,I type, which implies an age in the (10-30) Myr interval. However, using an empirical relation and integrated UBV colors, \citet{els85} derived a slightly larger value of 40 Myr. According to Bica et al. (1999, hereafter B99), SL\,106 could be physically connected to the giant interstellar shell LMC7 (SGshell-LMC7). More recently, fitting theoretical isochrones to their resolved color-magnitude diagrams, Glatt et al. (2010, hereafter GGK10) reported $E(B-V)$ = 0.05 and an age between 25 and 100 Myr for SL\,106. On the other hand, based on broadband photometry and applying different methods, \citet{pop12} and \citet{deg13} obtained nearly consistent results limiting the cluster age to the (20-85) Myr interval. In general, all the previously reported ages for SL\,106 are consistent with those derived in the current study (Table~\ref{tbl-5}). As shown in Fig. 3, the cluster observed spectrum compares very well with the YC.LMC template (12-40 Myr) taken from the spectral library of S95. The similarity between these two spectra is evident, as shown by the corresponding flux residual (Fig. 3). Fig. 5 shows the observed spectrum of SL106 and the synthetic spectrum fitted using the BC03 base. An age of 22 Myr and $E(B-V)$ = 0.08 was obtained through this match. Since the obtained residuals are smaller for the latter fit, we decided to assign heavier weight to the BC03 values (Table~\ref{tbl-5}).

\subsection{SL\,134}
\citet{luc70} identified this object as an OB association (LH\,12), while B96 classified it as a SWB\,I type cluster, which corresponds to the (10-30) Myr age range. SL\,134 has been described as a group of a few concentrated stars immersed in a faint nebulosity. B99 considered it to be a CN object, which means a stellar system with traces of emission. This is now confirmed by the emission lines clearly present in the observed spectrum of the cluster (Fig. 2). B99 suggested that SL\,134 could be physically related to BSDL 332, a cluster which, in turn, is possibly connected to the HII region LMC-DEM 39. As shown in Table~\ref{tbl-4}, both Balmer and metallic lines indicate age values lower than 50 Myr. From isochrone fitting, GGK10 derived $E(B-V)$ = 0.25 and a cluster age in the (25-158) Myr range. The observed integrated spectrum of SL\,134, corrected for $E(B-V)$ = 0.18, shows a reasonably good resemblance to the S95's YB.LMC template of 6-9 Myr (Fig. 3). The best fit with BC03 points to an age of 21 Myr and $E(B-V)$ = 0.19, as can be seen in Fig. 5. Although both fits are reasonably good, the emission lines suggest an age closer to that of the S95 template. This is why we adopted $E(B-V)$ = 0.18 and an age of 10 Myr for SL\,134 (Table~\ref{tbl-5}), in good agreement with the values determined by GGK10.

\subsection{SL\,142} 
This cluster is located in the outer parts of the LMC and, as far as we know, it has not been studied yet. From B96 we barely know that SL\,142 belongs to the SWB\,III type, which corresponds to the (70-200) Myr age range. Balmer lines suggest age values in the (100-500) Myr interval, while the SP04's diagnostic diagrams (DDs) indicate that SL\,142 lies in the limit that corresponds to the (40-350) Myr and (400-1500) Myr regions. As shown in Fig. 3, the spectrum of SL\,142, corrected for $E(B-V)$ = 0.05, compares reasonably well for $\lambda \ge$ 4000 \AA\, with the 50-110 Myr template from S95. Note, however, that for older ages, the S95 templates present a very limited spectral range ($\lambda \le$ 4500 \AA). Although the residuals obtained through both methods turn out to be rather poor, the fit obtained using BC03 base with a 242 Myr spectrum is clearly more accurate (Fig. 5). For this reason, we assign a heavier weight to the values determined using the BC03 base (Table~\ref{tbl-5}). These are the first age and reddening determinations for SL\,142.

\subsection{SL\,256} 
Also known as NGC 1848, this cluster was described by \citet{sha63} as a very compact group of stars. It belongs to the SWB\,I type (10-30 Myr) and, according to B99, it could be related to the OB association LH28. From isochrone fittings, GGK10 derived $E(B-V)$ = 0.15 and an age in the (32-126) Myr range. Both Balmer and metallic lines suggest that SL\,256 should be younger than 50 Myr, while the SP04's DDs indicate that it lies in the limit that corresponds to $<$ 40 Myr and (40-350 Myr) regions.  The observed integrated spectrum of SL\,256 resembles the S95's YC.LMC template of (12-40) Myr (Fig. 3), previously corrected for $E(B-V)$ = 0.15. A reasonable match is also found using the BC03 base with a 27 Myr synthetic spectrum, corrected by $E(B-V)$ = 0.18 (Fig. 5). The continuum distribution and depth of the Balmer lines are very similar in the two comparisons, so we adopted $E(B-V)$ = 0.16 and an age of 27 Myr for SL\,256 (Table~\ref{tbl-5}). The latter value is slightly lower than the one reported by GGK10.

\subsection{SL\,425}
This cluster, also classified by B96 as belonging to SWB\,I type (10-30 Myr), was described by \citet{sha63} as a fairly condensed group of stars. GGK10 determined a foreground reddening of $E(B-V)$ = 0.05 and an age in the (79-500) Myr range, while \citet{pop12} and \citet{deg13} reported for SL\,425 age ranges of (56-117) Myr and (52-144) Myr, respectively. 
The YC.LMC template of (12-40) Myr from S95's library, corrected for $E(B-V)$ = 0.25, yields the best match (Fig. 3), in accordance with the BC03 fitting (Table~\ref{tbl-5}) and with what the Balmer and metallic lines suggest (Table~\ref{tbl-4}). We adopted 27 Myr and the average $E(B-V)$ = 0.26 value for this cluster. 

\subsection{SL\,428}
This is a very young object also located in the outer parts of the LMC. B96 classified it as a SWB\,0 type (0-10 Myr) association embedded in a HII region. B99 reported that SL\,428 shows traces of emission because it is located in the HII region LMC-DEM155a. GGK10 determined for SL\,428 a colour excess $E(B-V)$ = 0.05 and an age in the (12-80) Myr range. The sum of the EWs of the Balmer and metallic lines suggest age values smaller than 40 Myr, while individual Balmer lines indicate ages between 10 and 500 Myr (Table~\ref{tbl-4}).  The observed integrated spectrum of SL\,428, with a reddening correction $E(B-V)$ = 0.05, compares reasonably well with the (3-6) Myr YA$_-$SG/WC.LMC template from S95 (Fig. 4). This template is representative of very young stellar populations in which hot supergiants seem to dominate the integrated light. Such template exhibits, in emission, the C\,III-IV (4650 \AA) spectral feature typical of Carbon Wolf-Rayet stars. The SL\,428 integrated spectrum does not show this emission feature but there is a noticeable similarity between those spectra, as can be seen n Fig. 4. An even better fit can be achieved using BC03 base (Fig. 6). According to this latter match, SL\,428 should be a practically unreddened 1 Myr old cluster. Therefore, this cluster seems to be slightly younger than was previously believed. 

\subsection{SL\,498}  
This is also a SWB\,0 type object (B96) located in the HII region LMC-DEM212 (B99). GGK10 obtained for SL\,498 $E(B-V)$ = 0.10 and estimated an age in the (19$\pm$80) Myr range, while \citet{pop12} and \citet{deg13} reported age ranges of (24-43) Myr and (12-16) Myr, respectively. As shown in Fig. 4, however, if the spectral region for $\lambda \le$ 4000 \AA\, is not taken into account, the best match is  found with the S95's YA.LMC template of (3-6) Myr, using $E(B-V)$ = 0.30. The match with the BC03 models turns out to be clearly better, even for $\lambda \le$ 4000 \AA\, (Fig. 6). Both these determinations reveal that we are dealing with a young cluster affected by a comparatively high reddening, for which reason we decided to assign heavier weight to the resulting values (Table~\ref{tbl-5}). Like SL\,428, this cluster seems to be somewhat younger than previously believed. Its comparatively higher reddening is possibly due to the fact that it is embedded in nebulosity.

\subsection{SL\,516}  
Like the two previous clusters, SL\,516 is a young SWB\,0 type object (B96). According to B99, it is associated with the HII region LMC-DEM214. Even though the S/N ratio of the cluster spectrum is rather low, the observed integrated spectrum of SL\,516 shows an acceptable resemblance to the S95's YA$_-$SG/WC.LMC template of 3-6 Myr (Fig. 4), once the former has been previously corrected for $E(B-V)$ = 0.15. The continuum distribution, the Balmer jump as well as the appearance and depth of some other spectral lines are nearly identical in both spectra, with the exception of the WC emission feature in 4650 \AA\,and the depth of H$\beta$. On the other hand, DDs, S$_h$ and S$_m$ indicate that SL\,516 should be younger than 40 Myr. This result agrees with the age of 5 Myr obtained from the fitting using BC03 base and $E(B-V)$ = 0.01. Since this match is better than the one obtained with the S95 template, we assigned heavier weight to the values inferred from using the BC03 base. GGK10 determined for SL\,516 an age range between 10 and 40 Myr and a color excess of $E(B-V)$ = 0.05, both values being slightly higher than those adopted here (Table~\ref{tbl-5}).

\subsection{SL\,543}  
This is a small and concentrated group of stars which could be physically connected with cluster BSDL332 (B99). According to B96, SL\,543 is a SWB\,III type cluster (70-200 Myr). GGK10 reported $E(B-V) = 0.11$ and an age between 40 and 160 Myr. EWs of metallic and Balmer lines suggest that SL\,543 should be between 10 and 350 Myr old.  The cluster spectral features are similar to those of the YDE.LMC (35-65 Myr) template from S95, provided that the observed spectrum is previously corrected by $E(B-V)$ = 0.11 (Fig. 4). These values are consistent with those determined by GGK10. On the other hand, the fit with the BC03 synthetic spectra suggests that this an almost unreddened 102 Myr old cluster. Both fits imply comparable values for age but not for reddening, being BC03 match clearly better. The finally adopted values are similar to those obtained from the synthetic spectra BaSTI base (Table~\ref{tbl-5}). 

\subsection{SL\,624}
As far as we know, no previous data exist for this object classified by B96 as belonging to the SWB type III (70-200 Myr). Therefore, the foreground reddening and age is here determined for the first time. 
The integrated spectrum of SL\,624, corrected for $E(B-V)$ = 0.10, exhibits good agreement with the YEF.LMC template of (50-110) Myr (Fig. 4). Note that this template covers a limited spectral range (see SL\,142). When the fittings using BC03 base were performed, we found that the residual fluxes are practically zero for the whole spectral range (Fig. 6). Consequently, we adopted an age of (110$\pm$30) Myr and $E(B-V)$ = 0.15. These values do not largely differ from those obtained with the S95 template (Table~\ref{tbl-5}).

\section{Concluding Remarks}
As part of a systematic spectroscopic survey of SCs in the LMC, we present in this paper flux-calibrated integrated spectra in the optical range for a sample of 10 blue concentrated SCs located in the inner disc and in the outer regions of the LMC. We derive here both foreground $E(B-V)$ reddening  and age values through different methods, using LMC template matching and fitting SSPs with two different models. Judging from the $\chi^2$ values, the matches between the observed and the synthetic spectra using SSPs from the BC03 base are noticeable better than those resulting from using the BaSTI base. SL\,142 and SL\,624 have not been previously studied so that the parameters determined here are the first of their kind. The derived parameters for the remaining 8 clusters show, in general terms, good agreement with those determined in previous studies. Cluster reddening values were also derived from available interstellar extinction maps, only for comparison purposes. With the exception of three clusters (SL\,134, SL\,425 and SL\,498), we find a satisfactory agreement between the $E(B-V)$ color excesses derived from template matching and the extinction maps, the mean difference (in absolute value) being $\Delta$$E(B-V)$ = 0.08 mag. The differences found for SL\,134, SL\,425 and SL\,498 may be related to the fact that at least two of them (SL\,134 and SL\,498) are clusters embedded in nebulosity. Three of the observed SCs (SL\,428, SL\,498 and SL\,516) turned out to be very younger than 5 Myr, whereas the remaining seven proved to be moderately young with ages between 10 and 240 Myr. The cluster spectra will later be used to generate new template spectra with the typical metallicity of the moderately young ($<$ 1 Gyr) LMC population. A star cluster spectral library at the LMC metallicity level can be useful for analyses of SCs in dwarf galaxies observable by means of ground-based large telescopes.





\acknowledgments
M.A.O and T.P. would like to thank the CASLEO staff members and technicians for their warm hospitality and support during the observing runs. We especially thank the referee for his valuable comments and suggestions, which helped us to improve the manuscript. This work was partially supported by the Argentinian institutions CONICET, FONCYT and SECYT (Universidad Nacional de C\'ordoba. Support for T.P. is provided by the Ministry of Economy, Development, and Tourism’s Millennium Science Initiative through grant IC120009, awarded to The Millennium Institute of Astrophysics, MAS. This research has made use of NASA's Astrophysics Data System Bibliographic Services. This research has made use of the SIMBAD database, operated at CDS, Strasbourg, France. This research has made use of the VizieR catalogue access tool, CDS, Strasbourg, France. The original description of the VizieR service was published in A\&AS 143, 23. This research has made use of ``Aladin sky atlas'' developed at CDS, Strasbourg Observatory, France (2000, A\&AS 143, 33 and 2014, ASPC 485, 277). The credits for the background image belongs to DSS used by ``Aladin sky atlas''.

\clearpage

\clearpage

\begin{figure}
\plotone{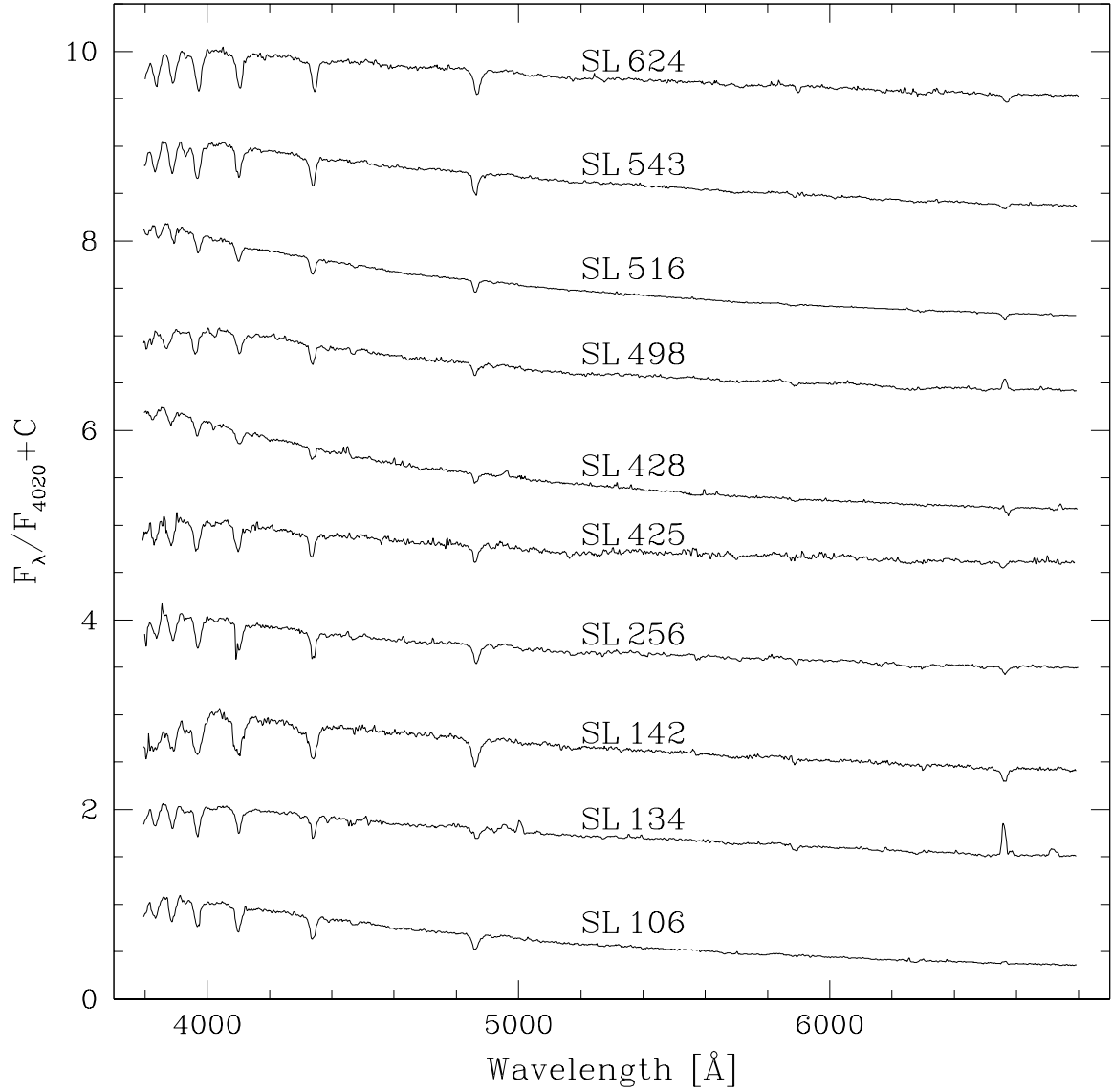}
\caption{ Observed integrated spectra of the whole SC sample. The spectra are in relative F$_{\lambda}$ units normalized at $\lambda$ $\approx$ 4020 \AA. Constants have been added to the spectra for clarity, except for the bottom one. \label{fig2}}
\end{figure}

\clearpage

\begin{figure}
\plotone{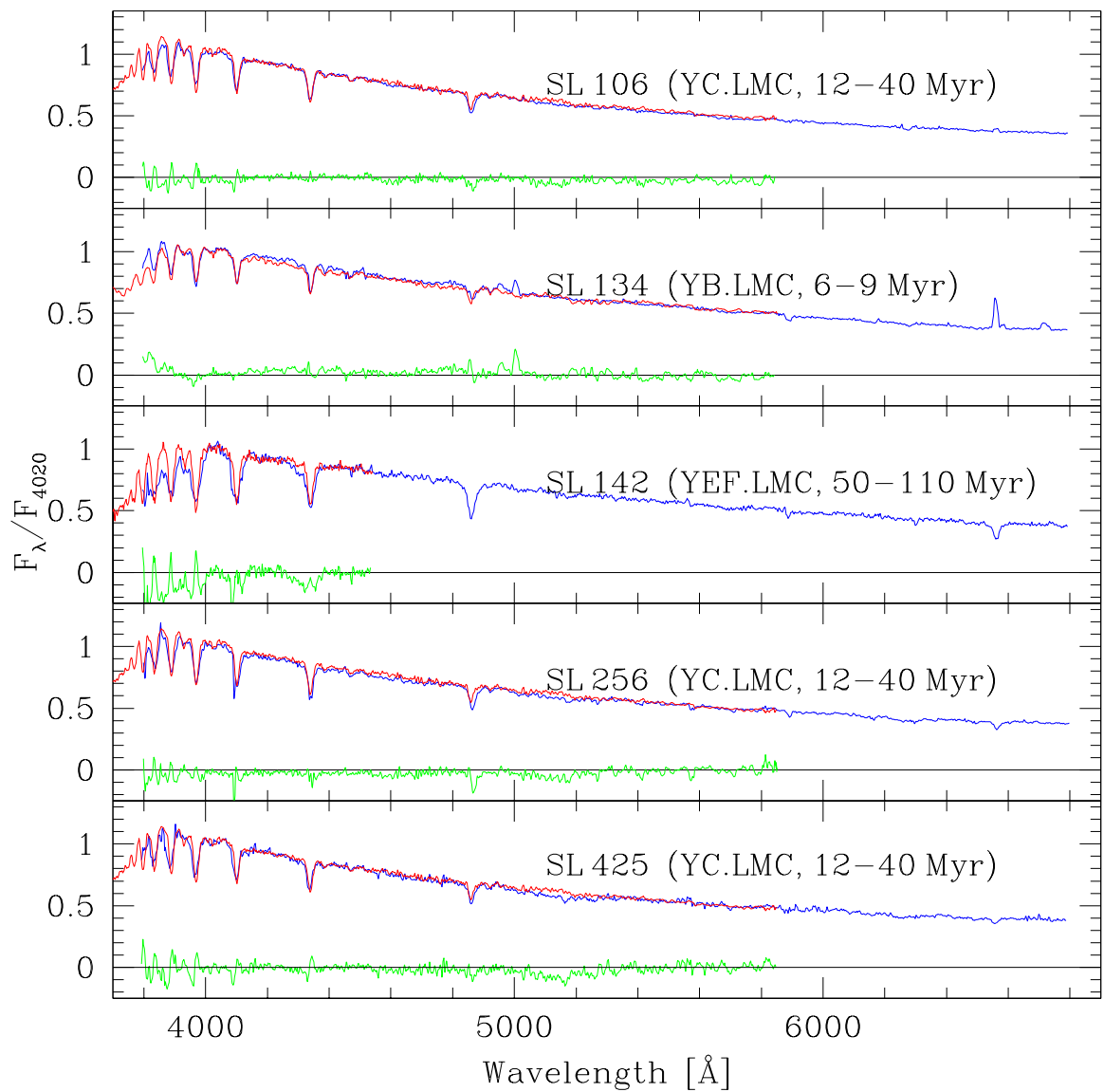}
\caption{ Results of the template matching method using S95 spectral library. Reddening corrected (blue), selected templates (red) and residual fluxes computed as (F$_{cluster}$ - F$_{template}$)/F$_{cluster}$ (green) are shown for SL\,106, SL\,134, SL\,142, SL\,256 and SL\,425. Units as in Fig. 2.} \label{fig3}
\end{figure}

\clearpage

\begin{figure}
\plotone{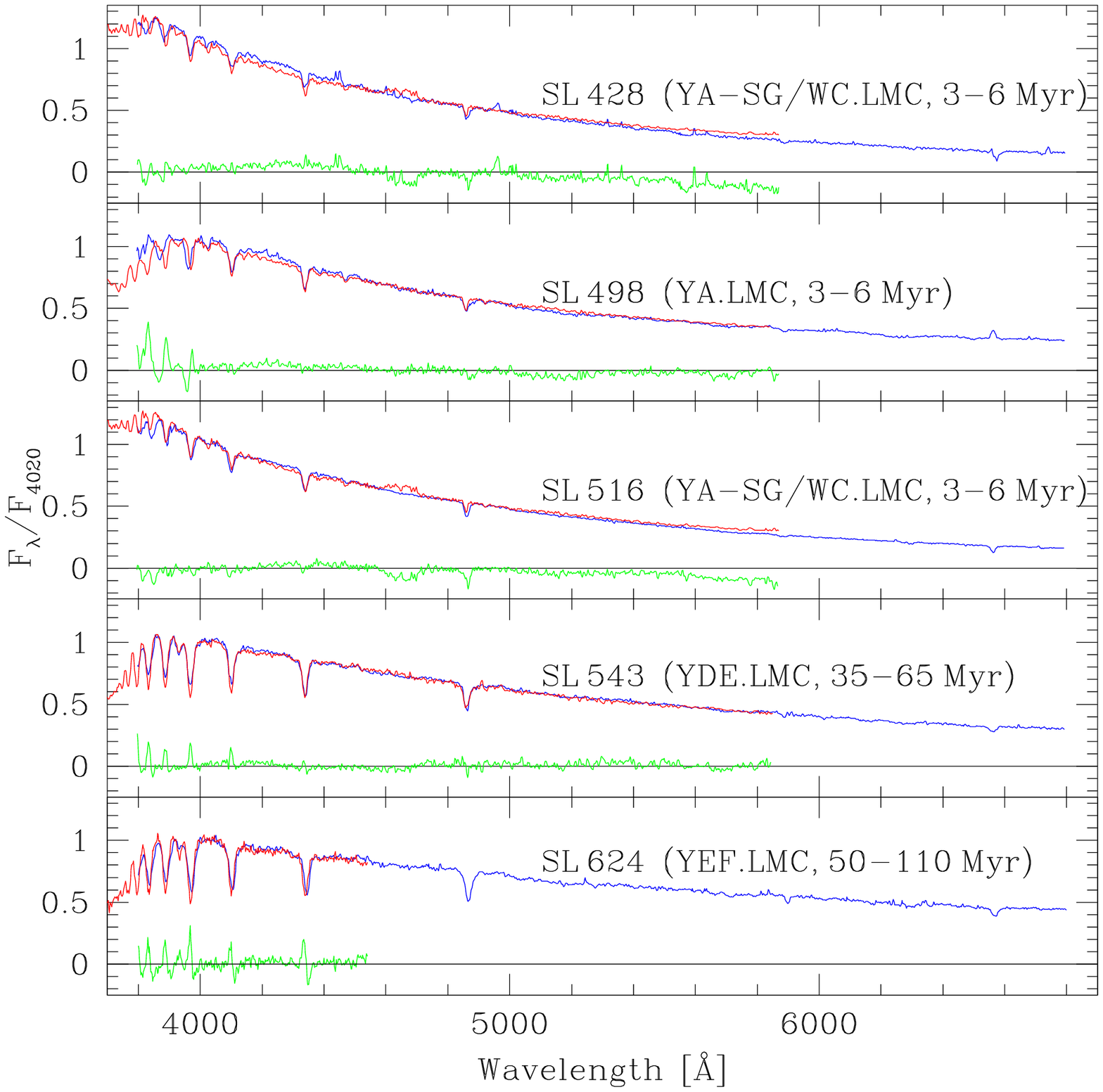}
\caption{Same as Fig. 3 for SL\,428, SL\,498, SL\,516, SL\,543 and SL\,624}.\label{fig4}
\end{figure}

\clearpage
\begin{figure}
\plotone{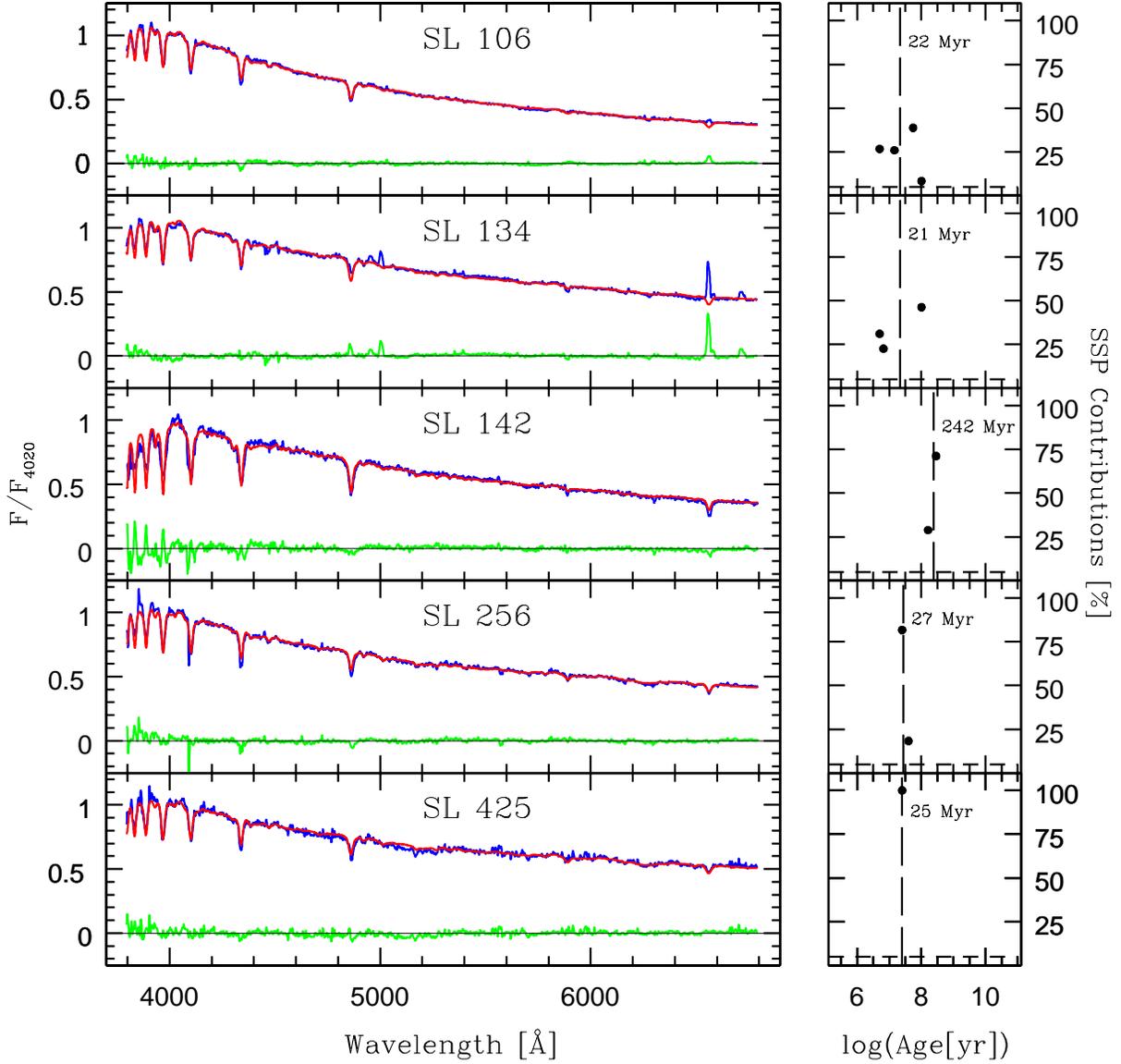}
\caption{{{\bf \sc starlight} results for clusters SL\,106, SL\,134, SL\,142, SL\,256 and SL\,425. Left panels: observed spectra (blue) and best fits of the spectra using BC03 models (red). A residual flux (green) is shown at the bottom of each panel, where a line at zero level has been drawn for reference. Right panels: results of the fits showing only the SSPs with non-zero contributions. Mean ages calculated calculated over the population vector of each fit are shown beside each vertical line. A horizontal dashed line has been drawn at 5\% level as a reference for negligible contributions to the total flux (see text for details). Units as in Fig. 2.}
\label{fig5}}
\end{figure}

\clearpage

\begin{figure}
\plotone{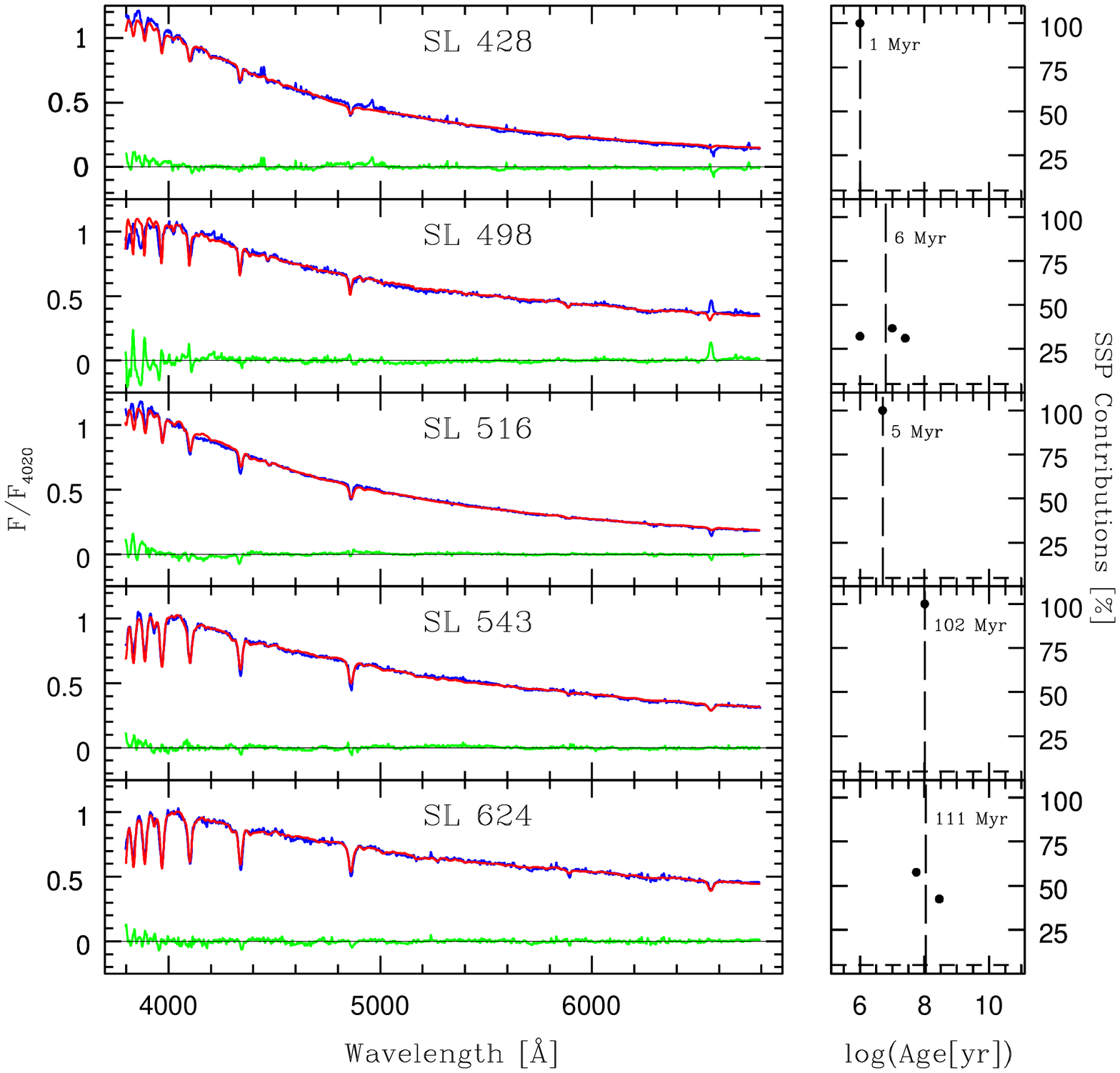}
\caption{Same as Fig. 5 for SL\,428, SL\,498, SL\,516, SL\,543 and SL\,624.} \label{fig6}
\end{figure}

\clearpage

\begin{figure}
\plotone{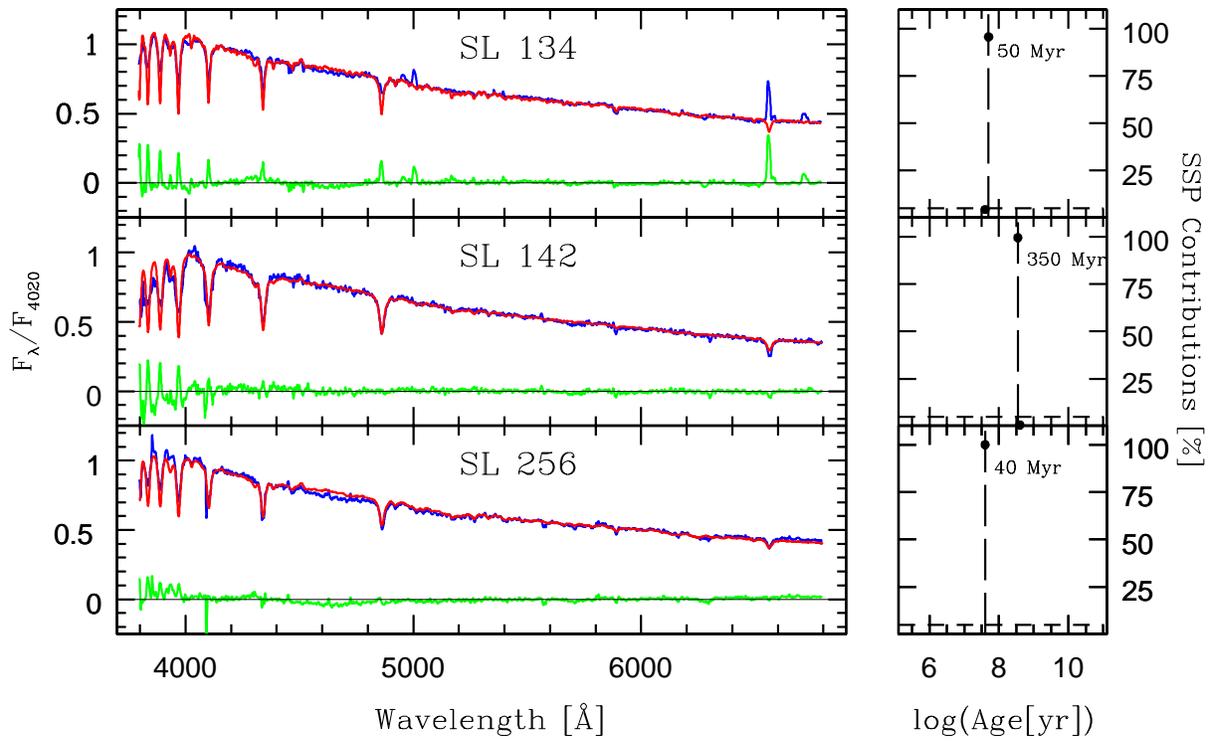}
\caption{Same as Fig. 5 for SL\,134, SL\,142 and SL\,256 but using BaSTI models.} \label{fig7}
\end{figure}

\clearpage

\begin{figure}
\plotone{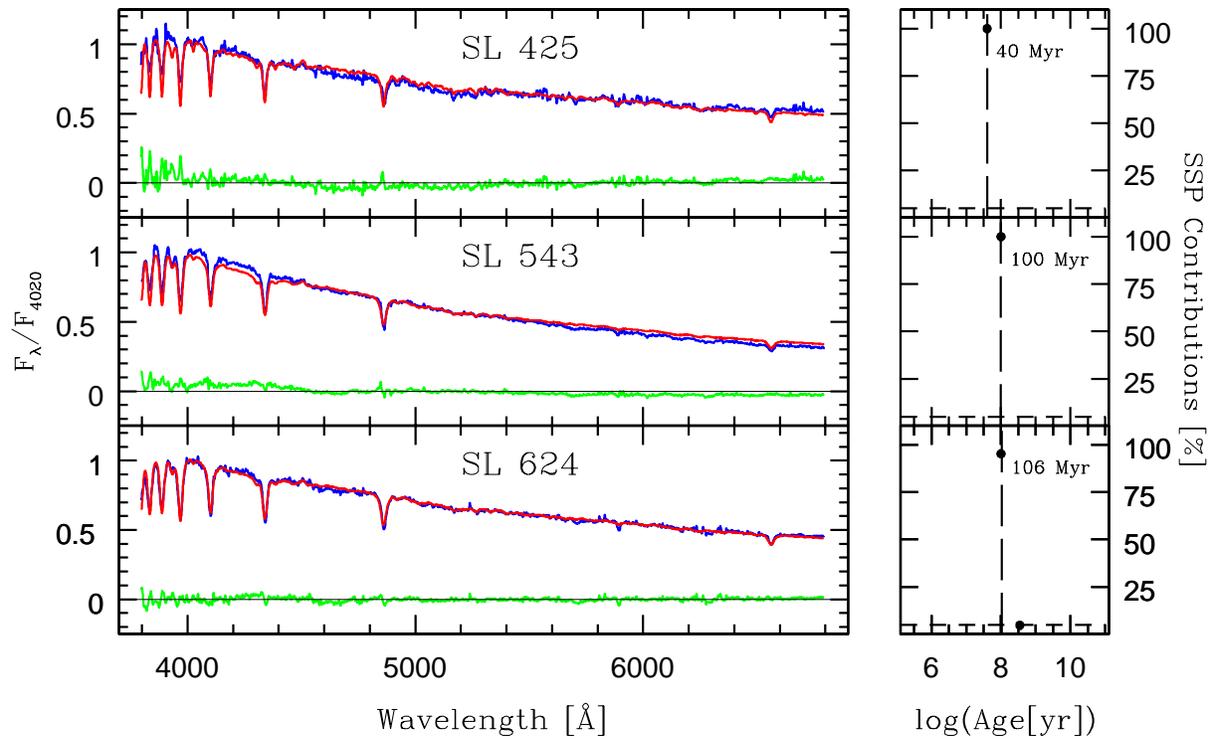}
\caption{Same as Fig. 7 for SL\,425, SL\,543 and SL\,624.} \label{fig8}
\end{figure}

\clearpage

\begin{figure}
\plotone{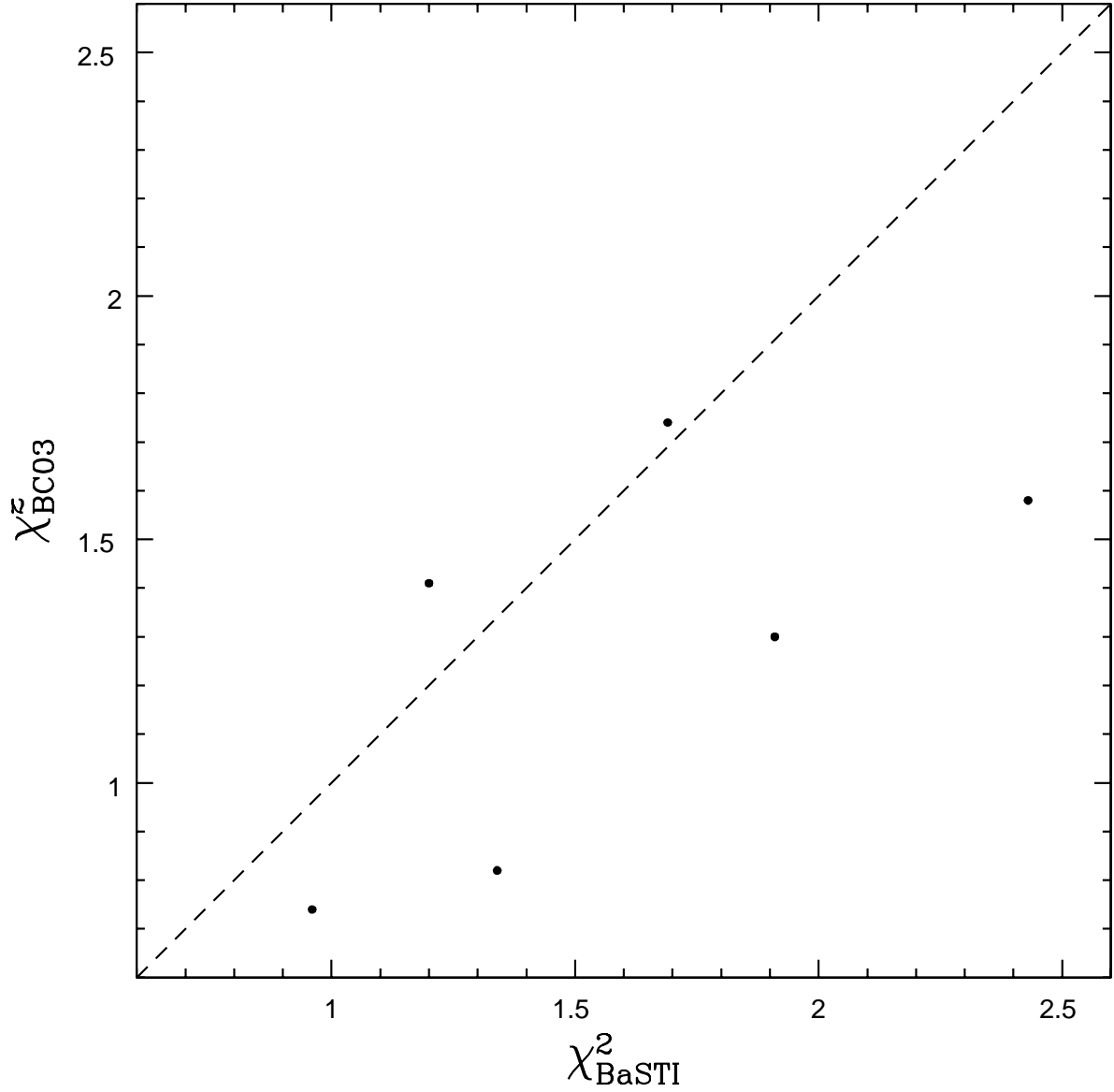}
\caption{Comparison of $\chi^2$ values of the fits obtained using BC03 and BaSTI models. The dashed line shows one-to-one correspondence. Fits performed using SSPs and BC03 models are clearly of better quality.}
\label{fig9}
\end{figure}

\clearpage

\begin{table}
\begin{center}
\caption{Cluster sample.\label{tbl-1}}
\begin{tabular}{lcccccc}
\tableline\tableline
Name\tablenotemark{a} & $\alpha_{2000.0}$ & $\delta_{2000.0}$ & {\it l} & {\it b} & D & SWB \\
                      &    (h m s)   & ($\degr\ \arcmin\ \arcsec$) & $\degr$  & $\degr$  & $\arcsec$ \\ 
\tableline
SL\,106, KMHK\,276 & 04 55 06 &  -69 40 24  & 281.095 & -35.301 & 34 & I \\  
SL\,134, KMHK\,349 & 04 57 30 &  -68 21 50  & 279.482 & -35.476 & 50 & I \\
SL\,142, KMHK\,357 & 04 58 15 &  -65 23 27  & 275.916 & -36.175 & 34 & III \\
SL\,256, NGC\,1848,  & 05 08 11 & -71 10 25 & 282.464 & -33.825 & 24 & I \\
KMHK\,580, ESO\,56-68&           &         & & & & \\
SL\,425              & 05 22 26 & -68 47 04  & 279.348 & -33.164 & 40 & I \\
SL\,428, KMHK\,833   & 05 22 51 & -65 43 00  & 275.713 & -33.616 & 100 & 0 \\
SL\,498, KMHK\,943   & 05 28 35 & -67 13 35  & 277.405 & -32.840 & 100 & 0 \\
SL\,516, KMHK\,972   & 05 30 00 & -66 58 40  & 277.090 & -32.735 & 50 & 0 \\
SL\,543, KMHK\,1016  & 05 30 60 & -71 53 34  & 282.824 & -31.905 & 50 & III \\
SL\,624, KMHK\,1158  & 05 37 19 & -71 06 59  & 281.830 & -31.532 & 50 & III \\
\tableline
\end{tabular}
\tablenotetext{a}{Clusters' designations from \citet{sha63} (SL), \citet{lau82} (ESO) and \citet{kon90} (KMHK).}
\end{center}
\end{table}

\clearpage

\begin{table}
\begin{center}
\caption{Log of observations.\label{tbl-2}}
\begin{tabular}{lccc}
\tableline\tableline
Cluster & Date & Exposure\tablenotemark{a} & S/N\tablenotemark{b} \\
\tableline
\objectname{SL\,106} & Dec. 21, 2011 & 1$\times$1800 + 1$\times$270 & 42 \\
\objectname{SL\,134} & Dec. 21, 2011 & 1$\times$1800 + 1$\times$2400 + 2$\times$3000 & 85 \\
\objectname{SL\,142} & Dec. 28, 2013 & 3$\times$3600 & 30 \\
        & Dec. 29, 2013 & 1$\times$3600 & \\
\objectname{SL\,256} & Dec. 20, 2011 & 2$\times$2400 & 35 \\
\objectname{SL\,425} & Dec. 21, 2011 & 1$\times$1800 + 1$\times$2700 & 45 \\
        & Dec. 28, 2013 & 2$\times$3600 & \\
\objectname{SL\,428} & Dec. 30, 2013 & 2$\times$3600 & 32 \\
\objectname{SL\,498} & Dec. 30, 2013 & 2$\times$3600 & 34 \\
\objectname{SL\,516} & Dec. 29, 2013 & 1$\times$1800 + 1$\times$2700 + 1$\times$4800 & 28 \\
\objectname{SL\,543} & Dec. 20, 2011 & 1$\times$2400 + 1$\times$3600 & 31 \\
\objectname{SL\,624} & Dec. 21, 2011 & 1$\times$2400 & 43 \\
       & Dec. 21, 2011 & 1$\times$3600 & \\
\tableline
\end{tabular}
\tablenotetext{a}{{\bf Exposure times in seconds}}
\tablenotetext{b}{S/N was measured in the (5000-5300 \AA) spectral range in the final combined spectra}
\end{center}
\end{table}

\clearpage

\begin{table}
\begin{center}
\caption{Equivalent widths (\AA)\tablenotemark{*}.\label{tbl-3}}
\begin{tabular}{lcccccccccc}
\tableline\tableline
Cluster & H$\alpha$ & Mg\,I & H$\beta$ & H$\gamma$ & CH G\,band & H$\delta$ & Ca\,II H & Ca\,II K & S$_h$\tablenotemark{1} & S$_m$\tablenotemark{2} \\
\tableline
Sl\,106	& -0.4 & -0.5 &	5.3	& 6.3 &	0.4 &	6.0 &	6.1 &	1.3 &	17.6 &	1.2 \\
Sl\,134	& -10.3 &	0.6 &	3.2 &	4.8 &	0.2 &	5.1 &	6.3 &	1.5 &	13.1 &	2.3 \\
Sl\,142	& 8.1 &	1.3 &	9.9 &	12.1 &	2.7 &	13.3 &	12.9 &	6.5 &	35.3 &	10.5 \\
Sl\,256	& 3.5 &	1.4 &	5.8 &	6.5 &	0.2 &	7.4 &	6.7 &	-0.4 &	19.7 &	1.2\\
Sl\,425	&2.5 &	1.7 &	4.6 &	5.9 &	0.7 &	6 &	6.4 &	-0.9 &	16.5 &	1.5 \\
Sl\,428	 & 6.2 &	0.1 &	3.5 &	3.2 &	0.2	& 3.3 &	3.1 &	0.3 &	10 &	0.6\\
Sl\,498	&-4.1&	1.2&	3.2&	5.1&	0.4&	4.5&	4.7&	0.4&	12.8&	2\\
Sl\,516	&5.2	&0.1&	4.2&	4.2&	-0.1&	4.1&	3.9&	0.9&	12.5&	0.9\\
Sl\,543	&3.4	&0.4&	6.8&	7.7&	0.6&	7.4&	8.7&	3.2&	21.9&	4.2\\
Sl\,624	&3.3	&0.4&	7.8&	8.7&	1.1&	9&	9.9&	1.9&	25.5&	3.4\\
\tableline
\end{tabular}
\tablenotetext{*}{Spectral windows were taken from \citet{bic86}: Ca\,II K = (3908-3952)\AA; Ca\,II H = (3952-3988); H$\delta$ = (4082-4124)\AA; CH G\,band = (4284-4318)\AA; H$\gamma$ = (4318-4364)\AA; H$\beta$ = (4846-4884)\AA; Mg\,I = (5156-5196)\AA; and H$\alpha$ = (6540-6586)\AA.}
\tablenotetext{1}{S$_h$ = H$\beta$+H$\gamma$+H$\delta$}
\tablenotetext{2}{S$_m$ = Mg\,I+CH G\,band+Ca\,II K}
\end{center}
\end{table}

\clearpage

\begin{table}
\begin{center}
\caption{Cluster ages from EWs\tablenotemark{*}.\label{tbl-4}}
\begin{tabular}{lcccc}
\tableline\tableline
Name & Balmer                & DD                   & S$_h$                   & S$_m$         \\
     & age\tablenotemark{1}  & age\tablenotemark{2} &    age\tablenotemark{2} &  age\tablenotemark{2} \\
\tableline
SL\,106 & 10-50 & $<$ 40 & 15 & 11 \\
SL\,134 & 0-50  & $<$ 40 & 8  & 17  \\
SL\,142 &100-500 & 40-350; 400-1500 & - & 290\\
SL\,256 & 10-50 & $<$ 40; 40-350 & 21 & 11 \\
SL\,425 & 10-50 & $<$ 40 & 12 & 13  \\
SL\,428 & 10-500 & $<$ 40 & 5 & 9 \\
SL\,498 & 0-50 & $<$ 40 & 8 & 15  \\
SL\,516 & 10-500 & $<$ 40 & 7 & 10 \\
SL\,543 & 10-100 & 40-350 & 32 & 36 \\
SL\,624 & 50-500 & 40-350 & 100 & 27\\
\tableline
\end{tabular}
\tablenotetext{*} {Age unit is Myr}
\tablenotetext{1}{According to \citet{bic86}}
\tablenotetext{2}{According to \citet{san06}}
\end{center}
\end{table}

\clearpage

\begin{table}
\begin{center}
\caption{Cluster parameters\tablenotemark{*}.\label{tbl-5}}
\begin{tabular}{lcccccccccc}
\tableline\tableline
 & BH82 & \multicolumn{2}{c}{S95's templates} & \multicolumn{2}{c}{BC03's SSPs} & \multicolumn{2}{c}{BaSTI's SSPs} & \multicolumn{2}{c}{Adopted values}   \\ 
Name & $E(B-V)$ & age & $E(B-V)$ & age & $E(B-V)$ & age & $E(B-V)$ & age & $E(B-V)$\\
\tableline
SL\,106 & 0.10 & 12-40 & 0.00   & 22   & 0.08     & -     & -    & 22$\pm$5   & 0.06$\pm$0.02 \\
SL\,134 & 0.05 & 6-9   & 0.18   & 21   & 0.19     & 50  & 0.05   & 10$\pm$5   & 0.18$\pm$0.02 \\
SL\,142 & 0.02 & 50-110& 0.05   & 242  & 0.00     & 350 & 0.00   & 240$\pm$70   & 0.00$\pm$0.02 \\
SL\,256 & 0.09 & 12-40 & 0.15   & 27   & 0.18     & 40    & 0.02 & 27$\pm$5   & 0.16$\pm$0.02 \\
SL\,425 & 0.06 & 12-40 & 0.25   & 25   & 0.27     & 40    & 0.11 & 27$\pm$5   & 0.26$\pm$0.02 \\
SL\,428 & 0.05 & 3-6   & 0.05   & 1    & 0.00     & -     & -    & 1$\pm$3    & 0.00$\pm$0.03 \\
SL\,498 & 0.06 & 3-6   & 0.30   & 6    & 0.18     & -     & -    & 5$\pm$2    & 0.20$\pm$0.10 \\
SL\,516 & 0.06 & 3-6   & 0.15   & 5    & 0.01     & -     & -    & 5$\pm$2    & 0.03$\pm$0.10 \\
SL\,543 & 0.08 & 35-65 & 0.11   & 102  & 0.01     & 100 & 0.00   & 100$\pm$50   & 0.01$\pm$0.06 \\
SL\,624 & 0.07 & 50-110& 0.10   & 111  & 0.15     & 106 & 0.11   & 110$\pm$30   & 0.15$\pm$0.05\\
\tableline
\end{tabular}
\tablenotetext{*} {Age unit is Myr}
\end{center}
\end{table}

\end{document}